\documentclass[11pt]{article}
\def\D{\Delta}
\def\d{\delta}
\def\L{\Lambda}
\def\l{\lambda}

\def\G{\Gamma}
\def\g{\gamma}
\def\e{\epsilon}

\def\o{\omega}
\def\i{\iota}
\def\a{\alpha}
\def\b{\beta}
\def\cd{\cal D}

\def\m{\mu}

\def\ha{\frac12}
\def\dim{\textrm{dim}}

\newcommand{\be}{\begin{equation}}
\newcommand{\ee}{\end{equation}}
\newcommand{\bea}{\begin{eqnarray}}
\newcommand{\eea}{\end{eqnarray}}

\begin{document}

\bigskip
\bigskip
\begin{center}
\bf{\Large Quantum Gravity as a Deformed Topological Quantum Field Theory}
\end{center}

\bigskip
\bigskip
\begin{center}
A. MIKOVI\'C\footnote{Talk presented at the QG05 conference, 12-16 September 2005, Cala Gonone, Italy.}\footnote{E-mail address: amikovic@ulusofona.pt}\\
Departamento de Matem\'atica,
Universidade Lus\'ofona de\\ Humanidades e Tecnologias,
Av. do Campo Grande, 376\\ 1749-024 Lisbon, Portugal
\end{center}

\bigskip
\bigskip
\centerline{{\bf Abstract}}

\bigskip
{\small It is known that the Einstein-Hilbert action with a positive cosmological constant can be represented as a perturbation of the $SO(4,1)$ BF theory by a symmetry-breaking term quadratic in the $B$ field. Introducing fermionic matter generates additional terms in the action which are polynomial in the tetrads and the spin connection. We describe how to construct the generating functional in the spin foam formalism for a generic BF theory when the sources for the $B$ and the gauge field are present. This functional can be used to obtain a path integral for General Relativity with matter as a perturbative series whose the lowest order term is a path integral for a topological gravity coupled to matter.}

\bigskip
\bigskip
\bigskip
Plebanski has shown that the General Relativity (GR) can be understood as a constrained
BF theory \cite{pleb}. This can be seen from the fact that the Einstein-Hilbert (EH) action in the Palatini formalism can be written as
\be S_{EH} = \int_M \e^{abcd}\,e_a \wedge e_b \wedge R_{cd} = \int_M B^{ab}R_{ab} \quad,\ee
where $e_a$ are the tetrads, $R_{ab} = d\o_a + \o_{ac}\wedge\o^c_a$ is the curvature two-form for the spin connection $\o_{ab}$ and the two-form $B$ is constrained by the relation 
\be B^{ab} =\e^{abcd} \,e_c \wedge e_d \quad.\label{esqr}\ee

Since the BF theories are topological, their quantization is easier, and hence one may try to quantize GR by imposing the constraints in the BF theory quantization scheme. This approach has been implemented by Barret and Crane in the spin foam formalism \cite{bce,bcl}, which is a discretized path-integral quantization. In this approach the spacetime manifold $M$ is replaced by a corresponding simplicial complex $T(M)$ (a triangulation of $M$) and the constraint (\ref{esqr}) becomes
\be \e_{abcd}\, B^{ab}_\D \,B^{cd}_\D = 0 \quad,\ee
where $B_\D =\int_\D B$ and $\D$ is a triangle of $T(M)$. This relation then leads to the construction of a state sum expression for the constrained BF theory path integral, which is of the spin foam type, i.e. one has a sum of the amplitudes for the labelled dual-two complex $\G (M)$ where the labels belong to a certain class of the unitary irreducible representations (irreps) of the Lorentz group.     

The BC approach gave a finite state sum for any non-degenerate triangulation of $M$ \cite{cpr}. However, the semi-classical limit of such a finite quantum gravity theory is
not known. Another problem is coupling of fermionic matter. The fermions couple to even and odd powers of the tetrads, while in the BC model one can couple the fermions only to powers of the $B$ field, which gives couplings only to the even powers of the tetrads. Although one can introduce the fermions by using the relation between a spin network and the corresponding spin foam \cite{amm},
this relation is not strong enough to determine the exact form of the amplitudes for a particle of a given spin. 

The problem with the fermions suggests to look for a BF theory formulation of GR where the tetrads are included. A natural way to do this is to consider the tetrads and the spin connection as parts of a larger connection, which then leads to the $SO(3,2)$ or $SO(4,1)$ BF theories. MacDowel and Mansouri have shown that the GR with a positive cosmological constant $\L_+$ can be understood as a broken symmetry phase of the topological Yang-Mills theory 
\be S_{TYM} = \int_M Tr\,\left( F \wedge F \right) \quad,\ee
where $F=dA +A\wedge A$ and $A$ is the $so(4,1)$ connection \cite{mdm}. Their results can be rewritten within the $SO(4,1)$ BF theory context as \cite{sm,sstd,frst}
\be S = \int_M Tr\,\left( {\bf B} \wedge F \right) - \frac{\a}{4}\, \int_M \e^{abcd}\,B_{ab}\wedge B_{cd} \quad,\label{bfe}\ee
where 
$$A=\o_{ab} J^{ab} + (2\L_+ )^{\ha}\,e_a P^a \,,$$
$$ F= (R_{ab} +\L_+ \,e_a\wedge e_b )J^{ab} + (de_a + \o_a^c\wedge e_c ) P^a \,,$$ 
$${\bf B} = B_{ab} J^{ab} + b_a P^a \quad.$$
$\a$ is the symmetry-breaking parameter, $J$ and $P$ denote the $so(4,1)$ algebra generators where $J$ denotes the $so(3,1)$ subalgebra generators. From the equations of motion one obtains  
$$B_{ab}=\frac{1}{2\a}\,\e_{abcd} (R^{cd} +2\L_+ \,e^c \wedge e^d ) \quad,$$
so that the remaining equations of motion are described by the action 
\be S' = \frac{\L_+}{\a}\int_M \e^{abcd}e_a\wedge e_b \left( R_{cd} + \L_+ \, e_c \wedge e_d \right) + 
\frac{1}{4\a}\int_M\e^{abcd}R_{ab} \wedge R_{cd} \,.\label{ehbf}\ee
The first term is the EH action with a cosmological constant, while the second term is the Euler class of $M$,
which does not affect the equations of motion. From (\ref{ehbf}) it is easy to see that $\a= G_N \L_+$, where $G_N$ is the Newton constant. If one takes the presently observed value for $\L_+$, this gives an extremally small value for $\a$, of the order of $10^{-120}$ \cite{frst}.

The smallnes of the dimensionless coupling constant $\a$ suggests that one could develop a perturbative quantization scheme of GR based on the $SO(4,1)$ BF theory. The path-integral for the action (\ref{bfe}) has a form
\be Z =\int {\cd}A {\cd}B \exp\left(i\int_M \,Tr\,(B\wedge F) + i\int_M d^4 x\,P(B)\right)\,,\ee
where $P(B)$ is a polynomial in $B$. If we couple the fermions, the total path-integral will require evaluating the path-integral of a more general form 
\be Z =\int {\cd} A {\cd} B \exp\left(i\int_M \,Tr\,(B\wedge F) + i\int_M d^4 x\,U(A,B)\right)\,,\label{papb}\ee
where $U(A,B)=P(B)+Q(A)$ and $Q(A)$ is a polynomial in $A$. For example, the Dirac fermion action can be written as
\be S_D = \int_M \e^{abcd}\,e_a \wedge \e_b \wedge e_c \wedge \bar\psi \left(\g_d D + m\, e_d \right)\psi  \quad,\ee
where $\g_a$ are the gamma matrices, $D=d + \ha\o^{ab}\, [\g_a ,\g_b ]$ is the spinorial covariant derivative and $m$ is the mass.

The path integral (\ref{papb}) can be evaluated perturbatively by using the generating functional technique. Let
\be Z_0 [j,J] =\int {\cd} A {\cd} B \exp\left(i\int_M \,Tr\,(B\wedge F) + i\int_M d^4 x\,Tr(jA + JB)\right)\,,\label{genf}\ee 
be the generating functional with the sources $j$ and $J$, then
\be Z=\exp\left( i\int_M d^4 x\,U\left(\frac{1}{i}\frac{\d}{\d j},\frac{1}{i}\frac{\d}{\d J}\right)\right)\,Z_0 [j,J]\Big{|}_{j=0,J=0}\,.\label{pertz}\ee

The generating functional (\ref{genf}) can be defined by using the spin foam formalism. Friedel and Krasnov have done this in the case when $j=0$ \cite{fk}, which would correspond to the case of pure gravity. We will describe here the general case. One begins by discretizing the spacetime manifold $M$, so that $M$ is replaced by a simplical complex $T(M)$. Then (\ref{genf}) can be written as
\be Z_0 [j,J] = \int \prod_l dA_l \prod_\D dB_\D \exp\left(i\sum_\D Tr\left(B_\D (F_f +J_f )\right) + i\sum_l Tr(j_l A_l )\right)\,,\ee
where $X_l =\int_l X$, $Y_f = \int_f Y$, $l$ is the dual complex edge and $f$ is the dual complex face. 

Instead of using the $A_l$ variables it is more convinient to use the holonomy variables
$g_l = e^{A_l}$, which are the group elements. The integral over the $B_\D$ variables has been defined in \cite{fk}, so that
\be Z_0 [j,J] = \int \prod_l dg_l \prod_f \d \left(g_{l_1 (f)}e^{J_{v_1 (f)}}g_{l_2 (f)}e^{J_{v_2(f)}}\,\cdots\,\right) \prod_l e^{iTr(j_l A_l )}\,,\ee
where $J_{v(f)}$ are portions of $J_f$ associated to each vertex $v$ of the face $f$ such that $J_{v_1(f)}+ J_{v_2 (f)} + \cdots = J_f$ and the $F_f$ is determined by $g_{l_1(f)} g_{l_2(f)}\cdots = e^{F_f}$. 

From the Peter-Weyl theorem we have
\be \d(g_f) = \sum_{\L_f} \dim \L_f \,Tr\, \left(D^{(\L_f )}(g_f)\right)\quad,\ee
and
\be e^{iTr(j_l A_l )} = \sum_{\l_l }\m^{\a_l}_{\b_l} (\l_l ,j_l )\, D^{(\l_l)\b_l}_{\a_l}(g_l) \quad,\ee
where
\be \m^{\a_l}_{\b_l} (\l_l ,j_l ) = \int_G dg_l \left(D^{(\l_l)\b_l}_{\a_l}(g_l )\right)^* e^{iTr(j_l A_l )}\,.\ee
We also need the formula
\be \int_G dg\,D^{(L_1)\b_1}_{\a_1}(g) \cdots D^{(L_n)\b_n}_{\a_n}(g) = \sum_\i C^{L_1 \cdots L_n (\i)}_{\a_1 \cdots \a_n}\left(C^{L_1 \cdots L_n (\i)}_{\b_1 \cdots \b_n}\right)^* \,,\ee
where $C^{(\i)}$ are the intertwiner coefficients for the tensor product of the irreps $L_1,\cdots,L_n$. By using these formulas we
arrive at
\be Z_0 [j,J] = \sum_{\L_f ,\l_l , \i_l} \prod_f \dim \L_f \Big{\langle} \prod_l \m (\l_l ,j_l ) \prod_v A_5 (\L_{f(v)}, \l_{l(v)},\i_{l(v)}, J_{f(v)}) \Big{\rangle} \quad.\label{sfgf}\ee

The novelty in the spin foam state sum (\ref{sfgf}) is the appearence of the tensorial weights $\m$ and $A_5$. The weight $\m$ carries two $\l$ indices, while the weight $A_5$ carry five $\l$ indices corresponding to five edges meeting at a vertex of the dual complex. The $\L$ dependence of $A_5$ comes from the fact that ten faces meet at a vertex of the dual complex, so that $A_5$ is given by the evaluation of the pentagon spin network with five free edges and ten internal edges having the insertions $D^{(\L)}(e^J)$. The $\i$'s denote the intertwiners which label the vertices of this open spin network. 

The free indices of $\m$ and $A_5$ have to be contracted, which is indicated by $\langle\,,\,\rangle$ in (\ref{sfgf}). This contraction can be represented as the evaluation of a spin network $\tilde{\G}(\L,J;\l,j;\i)$ associated to the dual one complex graph $\G$. The graph of the spin network $\tilde{\G}$ is obtained by replacing each vertex of $\G$ by a pentagon graph. The edges of the vertex pentagons of $\tilde\G$ carry the labels $\L$ and have the insertions $D^{(\L)}(e^J)$, while the edges connecting the pentagons carry the labels $\l$ and have the insertions $\m(\l,j)$. The vertices of $\tilde\G$ carry the intertwiners $\i$.  

The spin foam state sum (\ref{sfgf}) is an infinite sum, because a Lie group has infinitely many inequivalent irreps. Hence one has to insure the convergence of the sum (\ref{sfgf}). The standard regularization is to replace the irreps of $G$ with the irreps of the quantum group $G_q$ where $q$ is a root of unity. This gives a finite state sum of the same form as (\ref{sfgf}) 
\be Z^{(reg)}_0 [j,J] = \sum_{\L_f ,\l_l , \i_l} \prod_f \dim_q \L_f \Big{\langle} \prod_l \m (\l_l ,j_l ) \prod_v A^{(q)}_5 (\L_{f(v)}, \l_{l(v)},\i_{l(v)}, J_{f(v)}) \Big{\rangle} \quad,\label{qsfgf}\ee
where all the relevant spin network evaluations are replaced by the quantum spin network evaluations. The new sum is finite because the set of finite-dimensional irreps of a quantum group at a root of unity is finite. If one wants to use the category of unitary irreps, then the quantum group regularization is not guaranteed to work in the case of non-compact groups. In that case one obtains infinitely many quantum group irreps at roots of unity \cite{stein}. Still, there are special values of $q$ in the $SO(3,1)$ case which can give convergent state sums \cite{bcl}. Since there is no general theory of unitary quantum group irreps in the non-compact case, one needs to study the $SO(4,1)$ case.

An alternative approach to quantum group regularization would be an implementation of the Faddev-Popov gauge fixing procedure for spin foams proposed in \cite{fpsf}. 

After the regularization one can evaluate the partition function perturbatively in $\a$ as
\be Z (M) =\sum_{k=0}^\infty \a^k \,Z_k (M)\quad.\ee
Note that $Z_0 (M)$ is a manifold invariant, because it corresponds to a pure BF theory.
This invariant gives the signature for the four-manifold $M$ \cite{cyk}. However, $Z_k (M)$ for $k\ge 1$ will be generically triangulation dependent, which means that they will not be manifold invariants. Still, it would be worthwile exploring the $Z_k (M)$ because they may lead to new manifold invariants.  

As far as the physics is concerned, the important quantity to evaluate is the effective action $S_{eff} (A,B)$. It can be defined by a Legandre transform of the generating functional $Z[j,J]$. In the spin foam approach one would obtain the triangulation dependent expressions. The triangulation dependence has to be removed if one wants to find out what is the smooth-manifold limit. This would require the analysis of triangulations with increasing number of simplices. The smooth-manifold limit would then determine the classical and the semi-classical limit of the quantum theory defined by the spin foam state sum (\ref{qsfgf}). 

Note that a deformed BF theory can be quantized by using the usual continuum quantization techniques \cite{contq}, which could then lead to an effective action.

\end{document}